\documentclass{vgtc}                          

\ifpdf
  \pdfoutput=1\relax                   
  \usepackage{graphicx}                
  \DeclareGraphicsExtensions{.pdf,.png,.jpg,.jpeg} 
\else
  \usepackage{graphicx}                
  \DeclareGraphicsExtensions{.eps}     
\fi%

\graphicspath{{figures/}{pictures/}{images/}{./}} 

\usepackage{microtype}                 
\PassOptionsToPackage{warn}{textcomp}  
\usepackage{textcomp}                  
\usepackage{mathptmx}                  
\usepackage{times}                     
\usepackage{cite}                      
\usepackage{tabu}                      
\usepackage{booktabs}                  

\onlineid{0}

\vgtccategory{Research}

\vgtcinsertpkg

\title{Surfacing Misconceptions Through Visualization Critique}

\author{Amy Rae Fox\thanks{e-mail: amyraefox@ucsd.edu} %
\and Taylor Jackson Scott \thanks{e-mail:tjscott@ucsd.edu} %
}
\affiliation{\scriptsize Department of Cognitive Science \\ University of California, San Diego}
        
\abstract{ 
Students of visualization come to formal education with an abundance of personal experience. However, one's exposure to graphics through media and education may not be sufficiently diverse to appreciate the nuance and complexity required to design and evaluate effective representations. While many introductory courses in visualization address best practices for visual encoding of data based on perceptual characteristics, as cognitive scientists, we place equal value on representational decisions based on communicative context: \textit{how the representation is intended to be used}. In this pedagogical activity, we aim to surface learners' preconceived notions about what makes a visualization effective. Here we describe the structure and context of an introductory-level visualization activity, how it might be conducted in individual or group settings, our experience with the common misconceptions the activity can reveal, and conclude with recommendations on how they might be addressed. 
}

\CCScatlist{ 
 information visualization, visualization activity, pedagogy of data visualization, misconceptions
}

\begin{document}



\maketitle

\section{Introduction} 

Before misconceptions can be corrected, they need to be identified. As instructors of introductory courses on Information Visualization, we are astonished when we learn that, despite our best efforts, students have failed to grasp fundamental concepts about the variety of purposes that visualizations might serve. Often, introductory courses in visualization address best practices for visual encoding of data based on perceptual characteristics. This is an effective starting point for visualization design. However, rankings of visual variables and perceptual discriminability do not tell the whole story of our field. When designing and evaluating representations, we want our students to deeply consider the \textit{context} of artifacts, specifically, with respect to: (1) the audience (their prior knowledge, beliefs and skills), (2) the task: what the reader is expected to \textit{do} with the representation, and (3) the communicative context: the designer's goal in communication, be it to inform, persuade, mislead, record, or educate. In our experience, it is common for novice learners to persist with narrow assumptions about ``good" visualizations being dense, easy to read, reflections of ``truth" for some set of data.

Such misconceptions are often implicit, largely unquestioned, and deeply held. They  arise as \textit{preconceived notions} rooted in everyday experience, \textit{conceptual misunderstandings} from inaccurate application of prior knowledge, and \textit{factual misconceptions} in the form of unchallenged beliefs. Best practices in science education suggest that to supplant misconceptions with correct conceptual models, the inaccurate ideas must be identified and confronted \cite{national1997science}. Here we present a pedagogical activity that prompts learners to externalize their ideas about what makes visualizations effective or ineffective by preparing a design critique. Learners are challenged to explore the context of a visualization artifact and make inferences as to its audience and intended purpose, before evaluating the effectiveness of its design. Importantly, students are directed to give evidence to support their explanations, thereby prompting them to explore the sources of prior knowledge. By prompting students to make these ideas explicit, instructors can develop targeted examples to counter the most common misconceptions in follow-up instruction.

\section{Activity Description}

\textit{Synopsis: Students are asked to find two visualizations from educational materials or news media. Using the context of where the graphics are located, students make inferences as to the designer's intended audience and purpose. Students then provide objective critique of the effectiveness of each visualization. }

\subsection{Preparation}
As this activity is designed to elicit prior knowledge of visualization concepts, it can be performed early in an instructional session, with little to no prior instruction.  In the Spring of 2020 this activity was conducted as the first assignment in an online, introductory course on Information Visualization in the Department of Cognitive Science at the University of California, San Diego. The assignment was released during Week 2 (of a 10 week session), after lectures and required readings on: the history and modern context of visualization as a discipline \cite{few2009now}, distributed cognition as a theoretical framework \cite{liu2008distributed}, information and representation \cite{sedig2016design}, and visual variables and encoding \cite{sedig2016design}.

\subsection{Learning Goals}
\begin {enumerate}
\item The student's attention is brought to the \textit{context} in which visualizations are embedded in communication media, such as news articles, textbooks, and scholarly literature. 
\item The student develops awareness of the decisions a designer must make when developing a graphic to support a communicative goal. 
\item The student makes explicit their criteria for assessing a representation as effective with respect to its communicative goal. 
\end{enumerate}

\subsection{Context}

\paragraph{Asynchronous Activity.}Structured as a graded assignment, we asked the 38 students in our course to complete the activity independently over the course of one week. We expected students would spend 1-2 hours on the assignment, including searching for and critiquing their chosen visualizations. Students posted their responses on a course discussion forum, and were encouraged to review and comment on the submissions by classmates. Two weeks later, students completed a follow-up assignment where they reflected on their initial critiques and posted a response describing how their understanding of effective visualization had evolved. We expect the follow-up assignment took students between 15-30 minutes of reflection and writing. 

\paragraph {Synchronous Activity} With minimal modification, the assignment can be conducted as an interactive, synchronous, activity. In this case, we recommend providing learners with a small corpus of example visualizations, curated by the facilitator to include graphics that range in efficacy along any dimensions the instructor wishes to reinforce based on the context of the instructional session and background knowledge of participants. Depending on the number of participants, it may be most effective to have students form small groups where the collectively construct a critique, which can then be shared and discussed with the class. In the case of live sessions, it is necessary for facilitators to anticipate the most common misconceptions learners are likely to surface, and have prepared strategic counter-examples to address them.

\subsection{Instructions}

\textit{The following instructions are provided to learners.}

\paragraph{Background.} The use of visualization is pervasive in media: explanatory diagrams in magazines and online articles, graphs describing the projected impact of a new state budget, new experimental data plotted against theoretical expectations, demographic information, and of course—information on current events, etc. In each case, the author of the visualization tries to convey a point of view by choosing which data to present, emphasizing some aspects of the data while minimizing others. The result of these decisions can vary widely, from informative and enlightening, to confusing, or misleading.

\paragraph{Requirements.} Select two visualizations from any source (print or electronic). For each visualization, consider its context in order to make a subjective judgement as to the designer's purpose in creating it.  Then, develop a critique: an objective assessment of how well the visualization functions with respect to its intended purpose. 

\begin{itemize}

\item You should aim to select one visualization you judge to be effective, and one you judge to be ineffective. 

\item You should find visualizations ``in the wild", rather than texts or blogs on information visualization, where the graphic has already been analyzed and/or critiqued.  

\item In your description of each visualization, address the questions: Who is the intended audience? What is the purpose?  What type of data is shown? How is this data represented? What was the goal of the designer in representing the data in this way?

\item In your critique of the each visualization, use the concepts covered in the course to date to evaluate how well you think the design of the graphic functions in relation to its purpose. Are the data encoded effectively? What is the message the reader will likely take away? How are cognitive and perceptual principles being applied (or violated)? 

\end{itemize}

\section{Evaluation: Misconceptions, Revealed}

We developed this activity as a formative, rather than summative, assessment, aimed at scaffolding learners in their exploration of the purposes visualizations might serve, and explanation of criteria on which they should be evaluated. Although we did provide grades on the assignment (based primarily on effort, completeness of explanation, and citations to relevant course literature), the most impactful feedback to learners was provided in the lesson that \textit{followed} the assignment, which directly addressed the most common misconceptions present in student critiques through (theoretical) reinforcement of concepts, and targeted examples. We conducted this as a remote, synchronous lecture in our online course, though in a live setting, this could be conducted as a ``debriefing" session, provided the facilitator has anticipated the most likely misconceptions. In the sections that follow, we describe the most common misconceptions revealed by our Spring 2020 course, and recommendations for how they can be addressed. 

\subsection{Good visualizations are (immediately) easy to understand.}

The most pervasive misconception we observed in student critiques was the idea that to be effective, a visualization must be immediately easy to understand. This intuition seems to follow directly from popular discourse about the purpose of visualizations being to ``show" or ``reveal" data, and the adage that, ``a picture is worth one thousand words". These ideas are evident in responses like: 

\begin{itemize}
    \item \textit{``The visualization is good because it is very intuitive and easy to understand."}
    \item \textit{``This visualization is good because it is easy to make sense of the data at a glance..."}
    \item \textit{``The external representation of data isn’t simple enough for someone to read and understand instantly ..."}
    \item \textit{``There are too many visual marks to encode the visualization in a short amount of time."}
    \item \textit{``I would consider this a bad visualization because at first glance there is just too much going on."}
\end {itemize}

While each of these statements might have been accurate with respect to the particular visualization the student was critiquing, in their formulation of criteria, the learners have revealed that their conceptual model of efficacy does not appropriately rely on identification of the intended audience and  task. At the heart of this misconception are false assumptions about general audiences and quick-reading tasks. Not all visualizations are intended to be read by novices, or the general public. Nor are all visualizations intended for the purpose of ``informing" or ``quick reading".  We want students to understand that a seemingly complex visualization, one that is not easy to understand at first glance, might be \textit{very effective} if the intended audience and task require some degree of expertise and specialized prior knowledge of the domain, or graphical formalism. 

To address this misconception, we recommending presenting learners with example visualizations coming from the same information domain, with different analytical purposes. In the domain of weather, for example, one might contrast: (1) a 10-day weather forecast designed to \textit{inform} about temperature and rain predictions, to help
\textit{the general public} make \textit{decisions} about upcoming activities, with (2) a meteorological visualization such as a spaghetti plot \cite{rautenhaus2017visualization}, designed to support \textit{ meteorologists} interpreting surface fronts for \textit{forecasting} of cyclones. The key concepts to reinforce are: (1) the ease of readability should be evaluated with respect to the prior knowledge of the intended audience, and (2) the encoding structure should be judged with respect to the computational requirements of the expected task. An appropriate theoretical text on this concept is Stephen Palmer's, ``Fundamental Aspects of Cognitive Representation" \cite{palmer1978fundamental}, and in particular, the explanation of figure 9.1 (page 263).

 \subsection{The intended reader for a visualization is people who are interested in [topic of the visualization].}

To evaluate the communicative efficacy of a visualization, we teach students they must consider the intended audience of the graphic. Making inferences about an intended audience however, proved to be difficult for our students, as we saw an over-emphasis on the \textit{topic} (or domain) of data.
For example:
\begin{itemize}
    \item Regarding a graphic depicting COVID-19 diagnosis rates: \textit{``The intended audience is anyone who wants information about the current situation regarding COVID-19 because this news article can be viewed online without a subscription."}  Although the student has correctly noticed that the article is available without subscription, they fail to note the content of the news article, which focused on the economic impact of COVID-19. In this case, the graph was being used to support the journalist's arguments about economic impact in different countries.
    \item Regarding a graph depicting causes of death worldwide, from https://ourworldindata.org/. \textit{``The intended audience would be the entire world as it pertains to everyone."} The student has confounded the potential audience to whom the graphic might be relevant, with the likely audience for whom the author designed the graphic. 
    
\end{itemize}

Conversely, in critiquing a geothermal map depicting the crustal thickness of areas of the moon, one student concluded that the graphic was \textit{``most likely intended for astronomers, [or] researchers."} In this assertion, the learner reveals they have considered not only the data domain of the visualization, but also the media in which it was embedded, making reasonable inferences about what sub-population of readers interested in the topic the designer likely had in mind when designing the visualization. 

In another example, a student makes a reasonable inference about the intended audience based on the terminology used in labelling a figure (from The Economist magazine), but disagrees with the design decision, suggesting instead that the graphic would be more effective if designed for a general audience. \textit{``From the presentation of the line graphs, it is evident that the intended audience is not made to be public-friendly but targeted towards a specific group of people that are familiar with economic terms (e.g. maximum interest), such as people in the business field. Although one aspect that this visual lacks and would have made the visualization more effective is that it should have been targeted to be understood by everyone."}

A sophisticated understanding was demonstrated by a student who noted that a chloropeth map of COVID-19 cases within Cuyahoga county: (1) appeared in a locally-distributed newspaper and (2) included district labels that would only be relevant to locals. The student  correctly inferred that the audience was highly targeted and likely aimed at helping county residents make informed decisions about social distancing and other precautions against the virus.     

To confront the misconception that the audience of a graphic is the population of readers interested in the data being depicted, in follow-up lessons we emphasized the importance of \textit{communicative context}: Where is the artifact located, and what can that location tell us about the intention of the designer?  It can be effective to draw  contrast between conceptual artifacts like diagrams used to teach a concept, and analytical artifacts like statistical graphics used to communicate results in a scholarly paper. In turn we contrast these with an example of a persuasive graphic, where decomposition of the encoding structures and choice of data to include (and disclude) directly supports the narrative structure of media in which the graphic is embedded. In each case we prompt students to reflect on how the communicative context serves to narrow the potential audience for whom the graphic needs to be effective. 

 \subsection{``The purpose of the visualization is to convey [the data values in the visualization]."}
 
 This common misconception is similar to the last, but subtly different in an important way. Much like novices might presume that the intended audience of a visualization is as broad as \textit{those interested in the topic}, they might similarly presume that the purpose of the visualization is to \textit{convey the data} in the graphic.  The important component here regards \textit{what about} the data the author wishes to convey.  Wainer \cite{wainer_understanding_1992} fruitfully distinguishes between ``levels of reading", where a first-order reading involves extracting the value of individual datum, and second-order readings involve observing the relationship between data points \textemdash perceiving trends. In our experience, novices often presume that for a visualization to be effective, it must readily afford first-order readings. 
 
 \begin{itemize}
     
     \item Regarding a geographic heatmap visualization of COVID-19 cases in the United States: 
     \textit{``If the graph had used different colors maybe the audience can decipher the information better instead of the immediate thought of this [area] is bad”. } Here, the student has presumed that in order to be effective, the graphic \textit{should} support a first-order reading, allowing the user to extract the precise number of cases in a particular area.  In fact, the the context of the news article would suggest that the purpose of the graphic was to reinforce the author's narrative that Florida was a ``hot mess" of COVID-19 cases, prior to Spring Break. 
     
 \end{itemize}
 
 One can confront this misconception by directly teaching the ``levels of reading" concept (see \cite{wainer_understanding_1992}). It is also useful to present students with examples of statistical graphics from news articles that are aimed at supporting a narrative (the graphic quickly tells a single story), vs. those that are designed as exploratory (numerous outstanding examples from the New York Times Graphics team) that do not tell a single story ``at a glance", but rather, afford multiple readings.

\section{Conclusion}
By conducting this activity during our Spring 2020 introductory course on Information Visualization, we successfully brought students attention to pervasive misconceptions they held about the relationship between the purpose, task, and audience of communicative graphics.  We scaffolded their exploration of visualizations in news and educational media to help them make reasonable inferences about designers' intentions, and prompted them to make explicit their understanding of criteria on which visualizations should be evaluated. This allowed us to directly challenge misconceptions in our debriefing instructional sessions. In a follow up assignment, we asked learners to revisit their responses and ``critique" their critiques, reflecting on how their understanding of visualization purpose had evolved.  We hope the structure of this activity, and description of common misconceptions, will be useful for fellow instructors teaching the nuanced fundamentals of how visualizations can succeed as communicative artifacts.


\bibliographystyle{abbrv-doi}

\bibliography{template}

\begin{thebibliography}{1}

\bibitem{few2009now}
S.~Few.
\newblock {\em Now You See It: Simple Visualization Techniques for Quantitative
  Analysis. Chapter 1: Information Visualization}.
\newblock Analytics Press, 2009.

\bibitem{liu2008distributed}
Z.~Liu, N.~Nersessian, and J.~Stasko.
\newblock Distributed cognition as a theoretical framework for information
  visualization.
\newblock {\em IEEE transactions on visualization and computer graphics},
  14(6):1173--1180, 2008.

\bibitem{national1997science}
{National Research Council}.
\newblock {\em Science teaching reconsidered: A handbook. Chapter 4:
  Misconceptions as Barriers to Understanding Science}.
\newblock National Academies Press, 1997.

\bibitem{palmer1978fundamental}
S.~Palmer.
\newblock Fundamental aspects of cognitive representation.
\newblock {\em Cognition and Categorization}, pp. 259--302, 1978.

\bibitem{rautenhaus2017visualization}
M.~Rautenhaus, M.~B{\"o}ttinger, S.~Siemen, R.~Hoffman, R.~M. Kirby,
  M.~Mirzargar, N.~R{\"o}ber, and R.~Westermann.
\newblock Visualization in meteorology—a survey of techniques and tools for
  data analysis tasks.
\newblock {\em IEEE Transactions on Visualization and Computer Graphics},
  24(12):3268--3296, 2017.

\bibitem{sedig2016design}
K.~Sedig and P.~Parsons.
\newblock Design of visualizations for human-information interaction: A
  pattern-based framework. chapter 3: Conceptual elements of a framework.
\newblock {\em Synthesis Lectures on Visualization}, 4(1):1--185, 2016.

\bibitem{wainer_understanding_1992}
H.~Wainer.
\newblock Understanding {Graphs} and {Tables}.
\newblock {\em Educational Researcher}, 21(1):14--23, Jan. 1992. doi: {{%
10\hspace{.1pt}\discretionary{.}{%
}{.}\hspace{.4pt}3102\discretionary{/}{%
}{/}0013189X021001014}}


\end{thebibliography}
\end{document}